\documentclass[
aps,
prd,
10pt,
notitlepage,
superscriptaddress,
nofootinbib,
numbers,
showpacs
]{revtex4-1}

\usepackage[utf8]{inputenc}
\usepackage[T1]{fontenc}
\usepackage{anyfontsize}

\usepackage{amsmath}
\usepackage{tensor}
\usepackage{amssymb}
\usepackage{amsfonts}
\usepackage{mathrsfs}
\usepackage{bm}
\usepackage{graphicx}
\usepackage[dvipsnames]{xcolor}

\usepackage{silence}
\usepackage{hyperref}
\hypersetup{
    colorlinks=true,
    citecolor=Purple,
    linkcolor=Purple,
    urlcolor=Purple,
    linktocpage=true,
    breaklinks=true
}

\usepackage[capitalize]{cleveref}

\begin{document}

\title{Topological Thermodynamics of Generalized Bardeen Black Hole}

\author{A. A. M. Silva}
\email{anderson.alves@fisica.ufc.br}
\affiliation{Universidade Federal do Cear\'a (UFC), Departamento de F\'isica,\\ Campus do Pici, Fortaleza - CE, C.P. 6030, 60455-760 - Brazil.}

\author{M. H. Mac\^{e}do}
\email{matheus.macedo@fisica.ufc.br}
\affiliation{Universidade Federal do Cear\'a (UFC), Departamento de F\'isica,\\ Campus do Pici, Fortaleza - CE, C.P. 6030, 60455-760 - Brazil.}

\author{R. R. Landim}
\email{renan@fisica.ufc.br}
\affiliation{Universidade Federal do Cear\'a (UFC), Departamento de F\'isica,\\ Campus do Pici, Fortaleza - CE, C.P. 6030, 60455-760 - Brazil.}

\begin{abstract}

Neves and Saa introduced a two-parameter spacetime that encompasses the Bardeen and Hayward geometries as special cases, together with a case whose $f(r)$ function has the Simpson–Visser form. In this work, we employ the off-shell generalized Helmholtz free energy method to investigate the thermodynamic properties of this family within a topological framework. We begin by constructing the associated vector field and analyzing its zeros—specifically their winding numbers—to classify the thermodynamic branches and identify critical points associated with changes in global stability. Within the thermodynamic setup adopted here, the Bardeen, Hayward, and Simpson--Visser-type cases do not define independent global topological classes: whenever the off-shell conditions allow for two regular zeros, their winding numbers are $w=+1$ and $w=-1$, resulting in a total topological number of $W=0$. Within the more recent universal thermodynamic-topology classification, this stable-to-unstable defect ordering corresponds to the previously established $W^{0+}$ class, whereas the Schwarzschild limit is consistent with the $W^{1-}$ class. The parameters $\alpha$ and $\beta$ are useful for determining the extremal bound, the critical radius, the inverse critical temperature, and the extent of the locally stable branch. In contrast, the Schwarzschild limit exhibits a monotonic curve for the inverse temperature and a single topological defect with $W=-1$. These results provide a unified analytical characterization of how these configurations modify the local thermodynamic structure while preserving the global topological class of cases where $a$ is finite.

\end{abstract}

\keywords{Bardeen Black Hole; Thermodynamics; Topology} 

\maketitle

\section{Introduction}

In 1916, Schwarzschild discovered a spherically symmetric solution to Einstein's equations. This result was later understood as describing black holes, i.e., objects characterized by the presence of an event horizon. Since then, black holes have been confirmed by different lines of observational evidence, including the recent detections of gravitational waves by LIGO and the first images of the supermassive black hole obtained by the Event Horizon Telescope \cite{Abbott_2016,EventHorizonTelescope:2019dse}. However, the Schwarzschild solution predicts the existence of an unavoidable singularity at the center of the object. The existence of a central singularity inside a black hole is a major difficulty in general relativity because physical quantities become divergent at this point, indicating a breakdown of the classical description provided by general relativity.

Sakharov and Gliner suggested that essential singularity, like the one at the center of the Schwarzschild black hole, might be avoided if the usual vacuum, characterized by a vanishing energy--momentum tensor, were replaced by a vacuum-like medium described by a de Sitter geometry \cite{1966JETP,1966JETPGliner}. Following this idea, a number of singularity-free spacetimes, commonly referred to as regular solutions, were developed. Among them, Bardeen introduced the first nonsingular black-hole model with a de Sitter core, in which the constant mass is replaced by a position-dependent function \cite{1968Bardeen}. Other constructions, such as the Simpson--Visser solution, exhibit a geometry that can describe either a regular black hole or a wormhole, depending on the choice of parameters \cite{Simpson:2018tsi,lima2022blackstringbouncetraversable}. Another notable regular solution was proposed by Dymnikova, based on a gravitational analogue of the Schwinger effect \cite{Macedo:2024dqb,Macedo:2025guc,Ahmed:2026vce,Estrada:2019qsu}.

Later, Ayón-Beato and García demonstrated that the Bardeen solution can be obtained from the coupling of Einstein gravity with nonlinear electrodynamics \cite{Ayon-Beato:2000mjt}. In 2006, Hayward proposed another regular solution, a static and spherically symmetric black hole with a de Sitter core \cite{Hayward:2005gi}. The geometry was constructed so as to incorporate a finite curvature scale near the origin. Since then, the Hayward solution has been extended to include charged and rotating configurations \cite{Frolov:2016pav,Amir:2015pja}, along with investigations of their thermodynamics and quasinormal modes \cite{Molina:2021hgx,Lin:2013ofa}. Both solutions describe spacetimes with a highly dense central region of de Sitter type, characterized by $G_{\mu\nu}=-\lambda g_{\mu\nu}$ with $\lambda>0$. Recently, a generalization of the Bardeen metric has been proposed, and it can be demonstrated that the Hayward spacetime arises from this metric for a particular choice of parameters \cite{DuttaRoy:2022ytr}.

Bekenstein proposed that black holes have an entropy associated with the area of the event horizon and that their energy is associated to the black hole mass \cite{Bekenstein:1972, Bekenstein:1973}. Later, Hawking showed that black holes emit thermal radiation with a temperature proportional to the surface gravity of the black hole \cite{Hawking:1975}. Since the interior region of a black hole cannot be directly observed,  one may ask whether thermodynamic processes  outside the event horizon can distinguish regular geometries from their singular counterparts.

The thermodynamic description of regular black holes can be complicated due to the presence of additional terms in the first law of thermodynamics, which in turn lead to complications in establishing correspondences between mechanical and thermodynamic quantities \cite{Zhang:2016}. In addition, some regular black holes cannot simultaneously satisfy the area law, $S \neq A/4$, and the relation $dS = dM/T$ \cite{Lan:2023, Lan:2020fmn}. More recently, the thermodynamics of black holes from a topological perspective has received considerable attention recently \cite{SHAHZAD2025100900,Hazarika:2023iwp,Bao_2026, SEKHMANI2025102079,NASHED2025139866}. In this approach, the stability of a regular black hole can be analyzed by the topological property of the considered spacetime. In particular, the winding number can be used to characterize different thermodynamic phases and their stability, as well as to identify phase transitions and changes in the sign of the heat capacity \cite{Wei:2019uqg, Wei:2021vdx, Zhang:2025cev, silva2025topologicalthermodynamicsblackholes, Wu:2025xxo}.

The thermodynamic-topology framework based on the generalized off-shell Helmholtz free energy has subsequently been applied to a broad variety of black-hole spacetimes. Studies of neutral and charged Lorentzian NUT-charged geometries showed that the introduction of an additional parameter does not necessarily generate a new thermodynamic topological class, although its influence can depend on the asymptotic structure and on the thermodynamic variables considered \cite{Wu_2023_NUT,Wu_2023_ChargedNUT}. In contrast, investigations of static and rotating accelerating black holes demonstrated that acceleration and the cosmological constant can modify the topological number relative to the corresponding nonaccelerating solutions \cite{Wu_2023_Accelerating,Liu_2025}. Applications to static multi-charge and rotating charged AdS black holes in gauged supergravity further revealed that charge configurations and rotation parameters may modify the thermodynamic topological structure and, in some cases, lead to temperature-dependent topological transitions \cite{Wu_2024,Zhu_2024}.

A further development of the thermodynamic-topology framework was the introduction of a universal classification based not only on the total topological number $W$, but also on the asymptotic behavior and ordering of the thermodynamic branches. Within this scheme, four distinct classes, $W^{1-}$, $W^{0+}$, $W^{0-}$, and $W^{1+}$, were identified \cite{Wei:2024gfz}. An important consequence is that the same total topological number does not necessarily imply the same thermodynamic topological class; in particular, $W^{0+}$ and $W^{0-}$ both have $W=0$, but differ in the ordering and stability of their black-hole branches. The applicability and subsequent extension of this classification have been investigated for rotating black holes, three-dimensional BTZ black holes, multiply rotating Kerr--AdS solutions, multi-charge AdS black holes, and ultraspinning Kerr--AdS geometries \cite{Zhu_2025,Wu_2025,Chen_2025,Ai_2025,Wu_2026,Tian_2026}, leading to the identification of additional classes and subclasses beyond the original universal scheme.

In this work, we investigate whether the finite-$a$ limits contained in the Neves--Saa family realize distinct global thermodynamic topological classes or share a common defect structure. The object of study in this work is therefore the full Neves--Saa family rather than an isolated Bardeen-type geometry. The Bardeen and Hayward black holes, together with a Simpson--Visser-type case, are investigated as particular regular limits obtained from specific choices of the parameters $\alpha$ and $\beta$. By Simpson--Visser-type, we refer to the choice $\alpha=1$ and $\beta=2$, for which the metric function $f(r)$ coincides with the lapse function of the Simpson--Visser spacetime, although the angular sector, and hence the full geometry, is different. This distinction is important because the aim of the present analysis is to identify which features of the thermodynamic-topological structure are shared by the family and which depend on the selected regular black-hole realization.

 Our analysis is not intended merely to reproduce the heat-capacity behavior of the individual geometries. Instead, we derive a general analytical expression for the critical radius as a function of $\alpha$ and $\beta$, which allows us to distinguish the global topological number from the parameter-dependent locations and thermodynamic properties of the defects. We find that the Bardeen, Hayward, and Simpson--Visser-type cases considered here share the global value $W=0$ whenever two regular defects exist, whereas $\alpha$ and $\beta$ determine the extremal bound, the critical radius, the critical inverse temperature, and the domain of local stability. The Schwarzschild limit is qualitatively different and belongs to the $W=-1$ class. Therefore, the main physical contribution of the present work is a unified analytical characterization of how the parameters continuously modify the local thermodynamic structure without generating distinct global classes among the three finite-$a$ cases analyzed. In the universal classification of Ref.~\cite{Wei:2024gfz}, the $[+,-]$ ordering found for the Bardeen, Hayward, and Simpson--Visser-type cases corresponds to the $W^{0+}$ class, while the Schwarzschild limit is consistent with $W^{1-}$.

The paper is organized as follows. In Sec.~II, we present the generalized metric and its main thermodynamic properties. In Sec.~III, we develop the thermodynamic-topology analysis and discuss the Bardeen, Hayward, and Simpson--Visser case limits. In Sec.~IV, we present and compare the results, and in Sec.~V, we summarize our conclusions.

\section{Review of Generalized Bardeen Black Hole}

\begin{figure}[h!]
    \centering
    \includegraphics[width=0.75\linewidth]{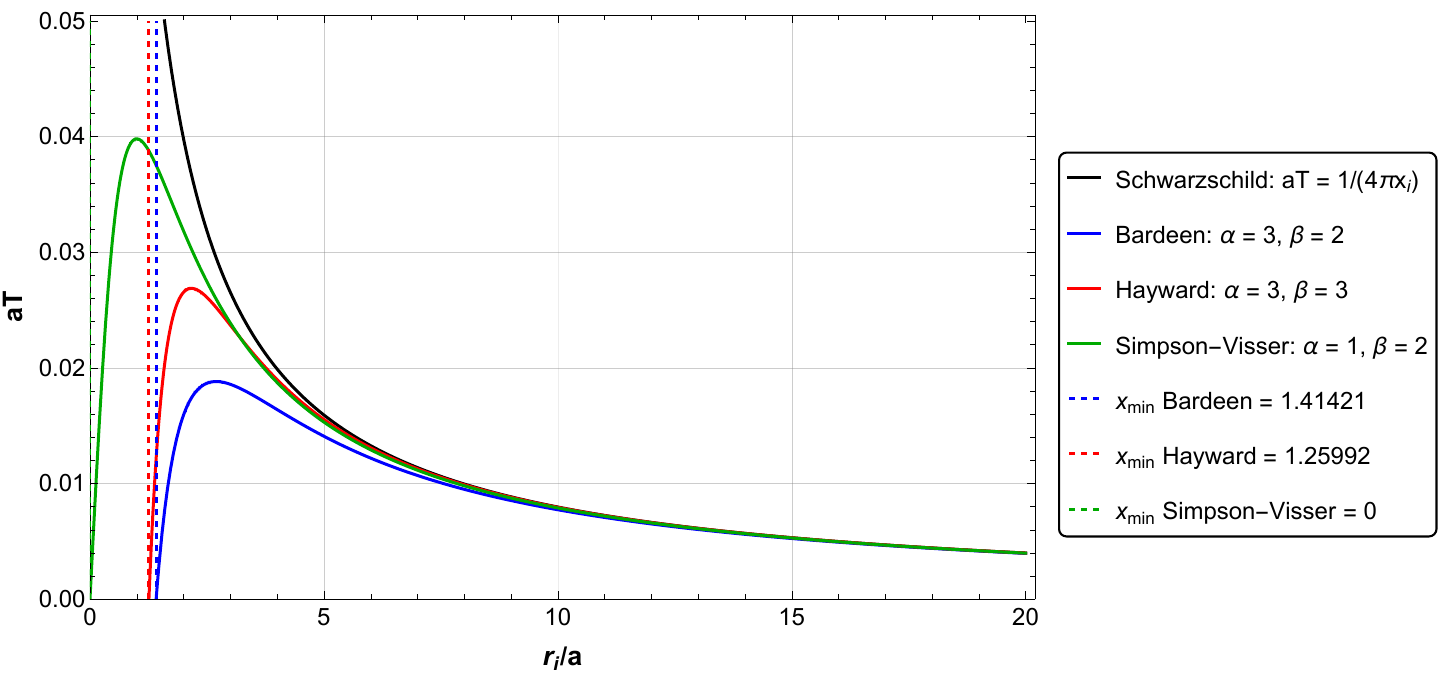}
    \caption{Plot of $aT$ as a function of $r_i/a$.}
    \label{fig:Figure 4}
\end{figure}

Neves and Saa obtained a static and spherically symmetric spacetime characterized by two classes of parameters which contains the Bardeen and Hayward metric as special cases, together with Simpson--Visser-type case \cite{Neves:2014aba}. This framework allows different regularization mechanisms to be analyzed in terms of their effects on the spacetime structure, mainly near the central region. The generalized Bardeen black hole has a line element of the form \cite{DuttaRoy:2022ytr}
\begin{equation}\label{2.1}
    ds^2 = -f(r)dt^2 + \dfrac{dr^2}{f(r)} + r^2d\Omega^2,
\end{equation}
where 
\begin{equation}\label{2.2}
    f(r) = 1 - \dfrac{2M r^{\alpha - 1}}{\left(r^{\beta} + a^{\beta}  \right)^{\alpha/\beta}}.
\end{equation}
Here $M$ and $a$ are the mass and length parameters, respectively, while $\alpha$ and $\beta$ are positive real parameters. This metric reduces to the  Bardeen metric when $\alpha =3$ and $\beta = 2$, to the Hayward metric if $\alpha = \beta = 3$ and to the Simpson--Visser-type  case defined in the introduction, provided that $M$ and $a$ are suitably identified. Furthermore, it approaches the Schwarzschild black hole in the limit $r\gg a$. A de Sitter-like behavior at small $r$ requires $\alpha = 3$, since
\begin{equation*}
    f(r ) \simeq  1 - \dfrac{2M r^{\alpha - 1}}{a^{\alpha}} 
\end{equation*}
which, for $\alpha = 3$, becomes
\begin{equation*}
    f(r) \simeq 1 - \dfrac{r^2}{r_0^2}, \qquad \qquad \text{where $r_0^2 = \dfrac{a^3}{2M}$}.
\end{equation*}
At large radial distances, $r \gg a$, the metric asymptotically approaches the Schwarzschild form,
\begin{equation*}
    f(r) \simeq 1 - \dfrac{2M}{r}.
\end{equation*}
 The black hole mass, as a function of the horizon radius, is obtained by setting $f(r_h) = 0$ and is given by
\begin{equation}\label{2.3}
    M(r_h) = \frac{1}{2} r_h^{1-\alpha } \left(r_h^{\beta } + a^{\beta }\right)^{\alpha /\beta }.
\end{equation}
Since $M(r_h)$ decreases for $r_h < r_{ext}$ and increases for $r_h > r_{ext}$, it possesses a global minimum at $r_{ext} = a(\alpha - 1)^{1/\beta}$. Therefore, horizons exist only for $M \geq M^{*} = M(r_{ext})$. For $M > M^{*}$, the equation admits two positive roots, corresponding to the inner and outer horizons, with $r_{-} < r_{ext} < r_{+}$. For $M = M^{*}$ these roots coincide. The  “remnant possibility” refers to the scenario where the Hawking temperature vanishes,  indicating a remnant black hole that does not evaporate completely. Black hole remnants have also been proposed as potential dark-matter candidates, since their stability and non-evaporating nature make them viable for such a role \cite{Chen:2002tu, Dymnikova:2010zz, Adler:2001vs}. For the generalized Bardeen metric, the Hawking temperature is
\begin{equation}\label{2.4}
    T = \dfrac{1}{4\pi}\dfrac{r_h^\beta - (\alpha - 1)a^\beta}{r_h(r_h^\beta + a^\beta)},
\end{equation}
which vanishes at $r_h = r_{ext}$. The corresponding mass is finite and given by $M^{*} = \frac{a\alpha^{\alpha/\beta}}{2(\alpha - 1)^{(\alpha - 1)/\beta}}$. Therefore, the extremal configuration can be interpreted as a black hole remnant.  It is worth noting that for $r\gg a$, the Hawking temperature reduces to that of a Schwarzschild black hole, while for $r \ll a$ and $\alpha = 3$ it approaches the de Sitter temperature. Figure~\ref{fig:Figure 4} shows the Hawking temperature as a function of the horizon radius. For all parameter choices considered here the temperature vanishes at $r_h = r_{ext}$. As $r_h$ increases, the Hawking temperature rises to a finite maximum and subsequently decreases, approaching the Schwarzschild black hole for $r_h \gg a$. 

\begin{figure}[h!]
    \centering
    \includegraphics[scale=0.53]{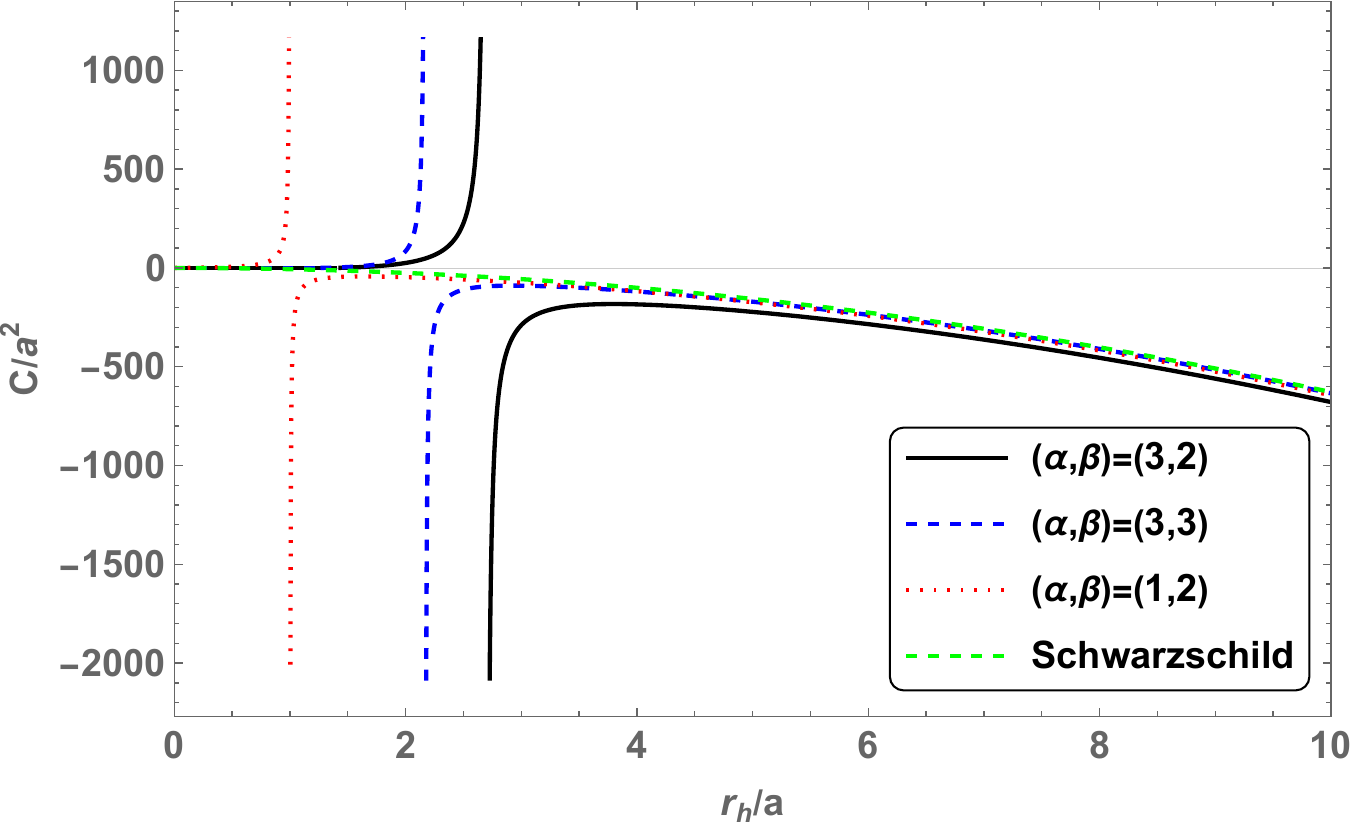}
    \caption{Heat capacity $C$ as a function of the dimensionless horizon radius for representative values of $\alpha$ and $\beta$. For $\alpha>1$, only the outer-horizon region $r_h/a>x_{\mathrm{ext}}=(\alpha-1)^{1/\beta}$ is physical; for the Simpson--Visser case ($\alpha=1$), the lower bound is $x_{\mathrm{ext}}=0$.}
    \label{Figure 2}
\end{figure}

A change in the sign of the heat capacity may indicate a possible phase transition in the black hole, whereas local stability is associated with a positive heat capacity while local instability occurs when the heat capacity is negative. The heat capacity of the generalized Bardeen black hole is given by:
\begin{equation}\label{2.5}
    C = \dfrac{2\pi r_h^{2-\alpha}(a^\beta + r_h^\beta)^{\frac{\alpha + \beta}{\beta}}[r_h^\beta - a^\beta(\alpha - 1)] }{(\alpha - 1)a^{2\beta} - r_h^{2\beta} + a^\beta r_h^\beta[\alpha(\beta + 1) - 2]   }.
\end{equation}
Figure \ref{Figure 2} displays plots of the heat capacity (\ref{2.5}) for different values of $\alpha, \beta$. The heat capacity has a single critical point \cite{Davies:1989ey}, which corresponds to the maximum of the Hawking temperature. Hence, the critical point determines the value of $r_{h}$ at which the generalized Bardeen black hole undergoes a phase transition. As we can see, the location of the critical point depends on $(\alpha, \beta)$.

The Kretschmann scalar invariant for metric (\ref{2.2}) has the form
\begin{equation}
    \begin{aligned}
    K & = 4M^2r^{2(\alpha -3)}(a^\beta + r^\beta)^{-2(2 + \frac{\alpha}{\beta})}\left\{12r^{4\beta} + a^{4\beta}[12 + \alpha(-20 + \alpha(17 -6\alpha + \alpha^2] - 4a^\beta r^{3\beta}[-12 + \alpha(5 + \beta)] \right. \\&
    \left. \qquad - 2a^{3\beta}r^{\beta}[-24 + \alpha(30 +2\beta - 17 \alpha - 3\alpha\beta + 3\alpha^2 + \alpha^2\beta) + a^{2\beta}r^{2\beta}(72 + \alpha(-60 - 8\beta + \alpha(17 + \beta(6 + \beta))\right\} .
\end{aligned}
\end{equation}
For $\beta > 0$ and $\alpha \geq 3$ the metric is regular everywhere, including at $r = 0$. It reduces to the Schwarzschild solution in the limit $a\rightarrow 0$ and approaches the de Sitter value $\frac{24}{r_0^4}$ when $\alpha = 3$ and $r \rightarrow 0$, acting as a limiting value for the spacetime curvature. Hence, the length parameter $a$ removes  the curvature singularity which occurs in the Schwarzschild black hole.

The components of Einstein equations yield
\begin{equation}
    \rho(r) = \dfrac{M\alpha a^\beta r^{\alpha - 3}}{4\pi(a^\beta + r^\beta)^{\alpha/\beta + 1}} = -p(r),
\end{equation}
\begin{equation}
    p_t = \dfrac{M\alpha a^\beta r^{\alpha - 3}}{8\pi (a^\beta + r^\beta)^{2 + \alpha/\beta}} [(1 + \beta)r^\beta -  a^\beta(\alpha - 1)].
\end{equation}

The null energy condition (NEC) is valid for all $r$ provided that $\alpha \leq 3$. For $\alpha > 1$, the strong energy condition (SEC) is violated near the center and is restored when  $p_t\geq 0$, which occurs for $r \geq \frac{r_{ext}}{(1 + \beta)^{1/\beta}}$.

\section{Topological Thermodynamic for Generalized Bardeen Black Hole}

\subsection{Thermodynamic Topology}

\subsubsection{General Case}

Let us now analyze the topological features of the generalized Bardeen black hole, following Refs.~\cite{Wei:2022dzw,Wei:2024gfz}. Before introducing the entropy, we specify the thermodynamic ensemble used throughout this work. The parameters $\alpha$ and $\beta$ are treated as fixed parameters labeling the different members of the spacetime family, and the regularization parameter $a$ is also held fixed during the thermodynamic variation. Therefore, the thermodynamic states considered here form a one-parameter family parametrized by the horizon radius $r_h$, with $da=d\alpha=d\beta=0$. Within this fixed-$a$ ensemble\footnote{ We emphasize that the above generalized off-shell Helmholtz free energy is defined for the fixed-$a$ ensemble adopted in this work. If the regularization parameter $a$ were promoted to a thermodynamic variable, the first law would in general contain an additional work term, schematically,
\begin{equation*}
    dM=T\,dS+\Psi_a\,da,
\end{equation*}
where $\Psi_a$ denotes the thermodynamic quantity conjugate to $a$. In such an extended thermodynamic phase space, the appropriate thermodynamic potential would depend on the chosen ensemble. Consequently, the off-shell vector field, its zeros, and the associated thermodynamic topological classification could in principle differ from those obtained here. We therefore do not claim ensemble independence of the topological number. The classification presented in this work applies specifically to the ensemble in which $a$, $\alpha$, and $\beta$ are held fixed}.
, the first law reduces to
\begin{equation}
    dM=T\,dS.
\end{equation}
More explicitly, writing $M = M(r_h)$, we have:
\begin{equation}
    dS=\frac{1}{T}\left(\frac{\partial M}{\partial r_h}\right)_{a,\alpha,\beta}dr_h.
\end{equation}
Thus, the entropy used below is defined along the fixed-$a$ thermodynamic branch as
\begin{equation}
    S(r_h)=\int\frac{1}{T}\left(\frac{\partial M}{\partial r_h}\right)_{a,\alpha,\beta}dr_h.
\end{equation}
Using Eqs.~(\ref{2.3}) and~(\ref{2.4}), one obtains
\begin{equation}
    \left(\frac{\partial M}{\partial r_h}\right)_{a,\alpha,\beta} = \frac{1}{2} r_h^{-\alpha}\left(r_h^\beta+a^\beta\right)^{\alpha/\beta-1}\left[r_h^\beta-(\alpha-1)a^\beta\right],
\end{equation}
and therefore
\begin{equation}
    \frac{1}{T}\left(\frac{\partial M}{\partial r_h}\right)_{a,\alpha,\beta} = 2\pi r_h^{1-\alpha}\left(r_h^\beta+a^\beta\right)^{\alpha/\beta}.
\end{equation}

The resulting entropy is then given by
\begin{equation}
    S(r_h) = \frac{2 \pi r_h^{2-\alpha}\left(a^{\beta}+r_h^{\beta}\right)^{\alpha/\beta}\left(a^{-\beta}r_h^{\beta}+1\right)^{-\alpha/\beta}\,{}_2F_1\left(\frac{2-\alpha}{\beta}, - \frac{\alpha}{\beta};\frac{2-\alpha}{\beta}+1;-a^{-\beta}r_h^{\beta}\right)}{2-\alpha},
\end{equation}
where $\,{}_2F_1\left(a,b;c;x\right)$ is the hypergeometric function. 
    
Using this entropy, we can compute the generalized off-shell Helmholtz free energy, defined as~\cite{Wu:2022whe,Wu:2023sue}
\begin{equation}
    \mathcal{F}=M-\frac{S}{\tau},
\end{equation}
where $\tau$ is the Euclidean time period, satisfying $\tau=1/T$ on shell. For the generalized Bardeen black hole, one obtains
\begin{equation}\label{generalized_free_energy}
    \begin{aligned}
        \mathcal{F} &= \frac{1}{2}r_h^{1-\alpha}\left(a^{\beta}+r_h^{\beta}\right)^{\alpha/\beta}\\
        &\quad - \frac{2 \pi r_h^{2-\alpha}\left(a^{\beta}+r_h^{\beta}\right)^{\alpha/\beta}\left(a^{-\beta}r_h^{\beta}+1\right)^{-\alpha/\beta}\,{}_2F_1\left(    \frac{2-\alpha}{\beta},-\frac{\alpha}{\beta};\frac{2-\alpha}{\beta}+1;-a^{-\beta}r_h^{\beta}\right)}{(2-\alpha)\tau}.
    \end{aligned}
\end{equation}

Following Ref.~\cite{Sadeghi:2023aii}, we define the vector field
\begin{equation}\label{vector_field_phi}
    \vec{\phi} = \left(\frac{\partial \mathcal{F}}{\partial r_h},-\cot\Theta\,\csc\Theta\right),
\end{equation}
where $0\leq \Theta\leq \pi$ is an auxiliary parameter. The zeros of this vector field are obtained from $\vec{\phi}=0$. Since the second component vanishes at $\Theta=\pi/2$, the first component gives
\begin{equation}
    \left.\frac{\partial \mathcal{F}}{\partial r_h}\right|_{r_h=r_i} = 0.
\end{equation}
From this condition, we obtain the Euclidean time period
\begin{equation}\label{tau}
    \tau = \frac{4\pi r_i\left(a^{\beta}+r_i^{\beta}\right)}{r_i^{\beta}+(1-\alpha)a^{\beta}}.
\end{equation}
This expression is the inverse of the Hawking temperature in Eq.~(\ref{2.4}), as expected from the on-shell relation $\tau=1/T$ in the topological thermodynamic approach~\cite{Wei:2022dzw}.

Introducing the dimensionless variable
\begin{equation}
    x_i=\frac{r_i}{a},
\end{equation}
and defining $\tau^*=\tau/a$, Eq.~(\ref{tau}) becomes
\begin{equation}\label{tau_dimensionless}
    \tau^* = \frac{4\pi x_i\left(1+x_i^\beta\right)}{x_i^\beta+1-\alpha}.
\end{equation}

For $\alpha>1$, the thermodynamic analysis must be restricted to the physical outer-horizon branch,
\begin{equation}\label{physical_outer_domain}
    \frac{r_i}{a}>x_{\mathrm{ext}}, \qquad x_{\mathrm{ext}}\equiv\frac{r_{\mathrm{ext}}}{a}=(\alpha-1)^{1/\beta}.
\end{equation}
At the endpoint $r_i=r_{\mathrm{ext}}$, the Hawking temperature vanishes and the on-shell Euclidean period diverges, $\tau\rightarrow\infty$. Consequently, this extremal configuration is the lower boundary of the physical branch and is not an isolated zero of the off-shell vector field for finite $\tau$. For the Simpson--Visser case, $\alpha=1$, the lower bound reduces to $x_{\mathrm{ext}}=0$, so that the physical branch is $r_i/a>0$.

In the asymptotic regime $r_i\gg a$, one has
\begin{equation}
    a^\beta+r_i^\beta \simeq r_i^\beta,\qquad r_i^\beta+(1-\alpha)a^\beta \simeq r_i^\beta.
\end{equation}
Therefore, Eq.~(\ref{tau}) reduces to
\begin{equation}
    \tau=4\pi r_i,
\end{equation}
which is precisely the Schwarzschild result obtained in Ref.~\cite{Wei:2022dzw}.

For the computation of the winding numbers, we follow the local topological analysis of Refs.~\cite{Wei:2022dzw,silva2025topologicalthermodynamicsblackholes}. For an isolated and nondegenerate zero of the vector field, the local winding number is determined by the orientation of the mapping $(r_h,\Theta)\mapsto(\phi^1,\phi^2)$, which is characterized by the Jacobian determinant
\begin{equation}\label{jacobian_definition}
    J_i = \left.\det\left[\frac{\partial(\phi^1,\phi^2)}{\partial(r_h,\Theta)}\right]\right|_{(r_i,\pi/2)},
\end{equation}
where
\begin{equation}
    \phi^1=\frac{\partial\mathcal{F}}{\partial r_h},\qquad\phi^2=-\cot\Theta\,\csc\Theta,
\end{equation}
so the Jacobian can be written explicitly as
\begin{equation}
    J_i = \left.\left[\frac{\partial\phi^1}{\partial r_h}\frac{\partial\phi^2}{\partial\Theta} - \frac{\partial\phi^1}{\partial\Theta}\frac{\partial\phi^2}{\partial r_h}\right]\right|_{(r_i,\pi/2)}.
\end{equation}
We note that since $\phi^1$ depends only on $r_h$ and $\phi^2$ depends only on $\Theta$, the mixed derivatives $\frac{\partial\phi^1}{\partial\Theta}$ and $\frac{\partial\phi^2}{\partial r_h}$ vanish. Furthermore, where $\vec{\phi} = 0$, we have $\Theta = \pi/2$, such that we obtain

\begin{equation*}
    \left.\frac{\partial\phi^2}{\partial\Theta}\right|_{\Theta=\pi/2} = \left.\frac{d}{d\Theta}\left(-\cot\Theta\,\csc\Theta\right)\right|_{\Theta=\pi/2} = \left.\frac{1+\cos^2\Theta}{\sin^3\Theta}\right|_{\Theta=\pi/2} =1.
\end{equation*}

Hence, for each regular zero,
\begin{equation}\label{jacobian_second_derivative}
    J_i = \left.\frac{\partial^2\mathcal{F}}{\partial r_h^2}\right|_{r_h=r_i}.
\end{equation}
It follows directly from the local orientation of the vector field that, for $J_i\neq0$, the winding number is
\begin{equation}\label{winding_numbers}
    w_i = \mathrm{sgn}(J_i) = \mathrm{sgn}\left[\left.\frac{\partial^2\mathcal{F}}{\partial r_h^2}\right|_{r_h=r_i}\right].
\end{equation}
Thus, the relation between the winding number and the second derivative of the off-shell free energy follows from the Jacobian of the vector-field mapping and is not inferred from the heat-capacity behavior. The degenerate case $J_i=0$ at the critical point will be discussed separately.

We now turn to the calculation of the winding numbers. For generalized free energy (\ref{generalized_free_energy}), we have
\begin{equation}
    \left.\frac{\partial^2\mathcal{F}}{\partial r_h^2}\right|_{r_h=r_i} = r_i^{-\alpha-1}\left(a^\beta+r_i^\beta\right)^{\frac{\alpha}{\beta}-2}\left[(\alpha-1)a^{2\beta} + a^\beta(\alpha\beta+\alpha-2)r_i^\beta - r_i^{2\beta}\right].
\end{equation}
Since $r_i>0$, $\alpha>0$, and $\beta>0$, the prefactor outside the square brackets is positive. Hence, the sign of the second derivative is determined only by analysis of the expression's sign 
\begin{equation*}
    G(r_i) = (\alpha-1)a^{2\beta}  + a^\beta(\alpha\beta+\alpha-2)r_i^\beta - r_i^{2\beta}.
\end{equation*}

Defining
\begin{equation}
    z_i=\left(\frac{r_i}{a}\right)^\beta,
\end{equation}
the sign analysis reduces to the quadratic function
\begin{equation}
    G(z_i) = (\alpha-1) + \left[\alpha(\beta+1)-2\right]z_i - z_i^2.
\end{equation}
Thus,
\begin{equation}
    w_i = \mathrm{sgn}\left[(\alpha-1) + \left[\alpha(\beta+1)-2\right]z_i - z_i^2\right].
\end{equation}

The roots of $G(z_i)$ are
\begin{equation}
    z_i = \frac{\alpha(\beta+1)- 2 \pm \sqrt{[\alpha(\beta+1)-2]^2+4(\alpha-1)}}{2}.
\end{equation}
For $\alpha\geq1$, the negative root is not physically relevant. Therefore, the only positive root is
\begin{equation}
    z_i^* = \frac{\alpha(\beta+1) - 2 + \sqrt{[\alpha(\beta+1)-2]^2+4(\alpha-1)}}{2}.
\end{equation}
Since $G(z_i)$ is a quadratic function with negative concavity, its sign is positive before the positive root and negative after it. Hence,
\begin{equation}
    \begin{cases}
        G(z_i)>0, & 0\leq z_i<z_i^*,\\[0.2cm]
        G(z_i)=0, & z_i=z_i^*,\\[0.2cm]
        G(z_i)<0, & z_i>z_i^*.
    \end{cases}
\end{equation}
Returning to the dimensionless variable $x_i=r_i/a$, we define the dimensionless critical radius
\begin{equation}\label{xcritical_general}
    x_c\equiv\frac{r_c}{a} = \left[\frac{\alpha(\beta+1) - 2 + \sqrt{[\alpha(\beta+1)-2]^2+4(\alpha-1)}}{2}\right]^{1/\beta}.
\end{equation}
For $\alpha>1$, the dimensionless extremal radius is
\begin{equation}
    x_{\mathrm{ext}}\equiv \frac{r_{\mathrm{ext}}}{a} = (\alpha-1)^{1/\beta}.
\end{equation}
In terms of the auxiliary variable $z=(r_i/a)^\beta$, the extremal point is therefore located at
\begin{equation}
    z_{\mathrm{ext}} = x_{\mathrm{ext}}^\beta = \alpha-1.
\end{equation}
Evaluating the quadratic function $G(z)$ at this point gives
\begin{equation}
    G(z_{\mathrm{ext}}) = \alpha\beta(\alpha-1)>0.
\end{equation}
Since $G(z)$ is a downward-opening quadratic with a unique positive root $z_c$, and $G(z)>0$ for $0\leq z<z_c$, the result $G(z_{\mathrm{ext}})>0$ implies $z_{\mathrm{ext}}<z_c$. Because the transformation $z=x^\beta$ is monotonic for $x>0$ and $\beta>0$, it follows that $x_{\mathrm{ext}}<x_c$.
Accordingly, the sign analysis must be restricted to the physical outer-horizon domain $r_i/a>x_{\mathrm{ext}}$. For $\alpha=1$, as in the Simpson--Visser case, $x_{\mathrm{ext}}=0$, and the physical domain begins at $r_i/a>0$.

For regular zeros, the radial classification is, therefore
\begin{equation}\label{winding_numbers_intervals_general_case}
    \begin{cases}
        w_i=+1, & x_{\mathrm{ext}} < \dfrac{r_i}{a}<x_c,\\[0.3cm]
        J_i = 0, & \dfrac{r_i}{a}=x_c, \quad \text{(degenerate zero)}\\[0.3cm] 
        w_i=-1, & \dfrac{r_i}{a}>x_c.
    \end{cases}
\end{equation}
The middle line must not be interpreted as $w_i=0$. The relation $w_i=\mathrm{sgn}(J_i)$ derived above applies only to isolated nondegenerate zeros with $J_i\neq0$. At $r_i/a=x_c$, one has
\begin{equation}
    \left.\frac{\partial\mathcal{F}}{\partial r_h}\right|_{r_i}=0,\qquad\left.\frac{\partial^2\mathcal{F}}{\partial r_h^2}\right|_{r_i}=0,
\end{equation}
so that $J_i=0$ and the critical configuration is a degenerate zero of the vector field.

The role of this degenerate point becomes transparent from the dimensionless inverse-temperature curve. Writing $x=r/a$ and using Eq.~(\ref{tau_dimensionless}), one finds
\begin{equation}\label{tau_derivative_G}
    \frac{d\tau^*}{dx} = -\frac{4\pi\,G(x^\beta)}{\left(x^\beta+1-\alpha\right)^2}.
\end{equation}
Since $G>0$ on the physical branch for $x_{\mathrm{ext}}<x<x_c$ and $G<0$ for $x>x_c$, the function $\tau^*(x)$ decreases up to $x_c$ and increases thereafter. Therefore,
\begin{equation}
    \tau_c^*\equiv\tau^*(x_c)
\end{equation}
is the minimum inverse temperature of the physical branch. Consequently, the defect structure depends on the control parameter $\tau^*$ as follows:
\begin{equation}\label{tau_defect_regimes}
    \begin{cases}
        \tau^*>\tau_c^*: & \text{two regular zeros }x_-<x_c<x_+,\quad w_- = +1,\; w_+=-1,\\[0.2cm]
        \tau^*=\tau_c^*: & \text{one degenerate zero at }x=x_c,\quad J_c=0,\\[0.2cm]
        \tau^*<\tau_c^*: & \text{no physical zeros of the off-shell vector field.}
    \end{cases}
\end{equation}
Near the critical point, the first derivative of $\tau^*$ vanishes and the local expansion is
\begin{equation}
    \tau^*(x)-\tau_c^* \simeq \frac{1}{2}\left.\frac{d^2\tau^*}{dx^2}\right|_{x_c}(x-x_c)^2,
\end{equation}
with $\left.d^2\tau^*/dx^2\right|_{x_c}>0$. Hence, for $\tau^*>\tau_c^*$, the two nearby zeros behave as
\begin{equation}
    x_{\pm}\simeq x_c \pm \sqrt{\frac{2(\tau^*-\tau_c^*)}{\left.d^2\tau^*/dx^2\right|_{x_c}}}.
\end{equation}
This expression explicitly shows the coalescence of the $w=+1$ and $w=-1$ defects as $\tau^*\rightarrow\tau_c^{*+}$. Their net charge is then
\begin{equation}
    \lim_{\tau^*\to\tau_c^{*+}}(w_-+w_+)=+1-1=0,
\end{equation}
but this vanishing total charge must not be confused with assigning an individual winding number $w=0$ to the degenerate zero itself.

Since the sign of the winding number agrees with the sign of the heat capacity away from the degenerate point, as discussed in Ref.~\cite{silva2025topologicalthermodynamicsblackholes}, one obtains

\begin{equation}\label{heat_capacity_intervals}
    \begin{cases}
        \mathrm{Sgn}(C)=+1, & x_{\mathrm{ext}}<\dfrac{r_i}{a}<x_c, \quad \text{topologically stable branch},\\[0.3cm]
        \mathrm{Sgn}(C)=-1, & \dfrac{r_i}{a}>x_c, \quad \text{topologically unstable branch}.
    \end{cases}
\end{equation}

At the degenerate point $r_i/a=x_c$, the condition
\begin{equation}
    M''S'-M'S'' = 0
\end{equation}
is satisfied. This is the critical configuration at which the heat capacity diverges or changes sign and the two regular topological defects coalesce.

In the asymptotic regime, $r\gg a$, the generalized metric approaches the Schwarzschild limit. Since this regime corresponds to $r/a>x_c$, one obtains $w_i=-1$ and therefore $\mathrm{Sgn}(C)=-1$, consistently with the well-known thermodynamic instability of the Schwarzschild black hole~\cite{Wei:2022dzw}.

\subsubsection{Bardeen Black Hole for $\alpha = 3$ and $\beta = 2$}

We now consider the topological structure of the Bardeen black hole, obtained from the generalized metric by taking $\alpha=3$ and $\beta=2$. In this case, the mass as a function of the horizon radius is given by
\begin{equation}\label{bardeen_mass}
    M_{\mathrm{Bardeen}} = \frac{\left(a^2+r_h^2\right)^{3/2}}{2r_h^2}.
\end{equation}
The corresponding Hawking temperature is
\begin{equation}\label{temperature_bardeen}
    T_{\mathrm{Bardeen}} = \frac{r_h^2-2a^2}{4\pi r_h\left(r_h^2+a^2\right)}.
\end{equation}
The entropy obtained from $S=\int dM/T$ reads
\begin{equation}\label{entropy_bardeen}
    S_{\mathrm{Bardeen}} = \frac{\pi\left(r_h^2-2a^2\right)\sqrt{a^2+r_h^2}}{r_h} + 3\pi a^2\tanh^{-1}\left(\frac{r_h}{\sqrt{a^2+r_h^2}}\right).
\end{equation}
These expressions are consistent with the results reported in the literature~\cite{Sharif:2010pj}. From these quantities, the generalized off-shell Helmholtz free energy is obtained as
\begin{equation}\label{generalized_free_energy_bardeen}
    \mathcal{F}_{\mathrm{Bardeen}} = \frac{\sqrt{a^2+r_h^2}\left[a^2(4\pi r_h+\tau) + r_h^2(\tau-2\pi r_h)\right] - 6\pi a^2 r_h^2\tanh^{-1}\left(\frac{r_h}{\sqrt{a^2+r_h^2}}\right)}{2r_h^2\tau}.
\end{equation}

The zeros of the vector field defined in Eq.~(\ref{vector_field_phi}) are obtained and we obtain for the parameter $\tau$ for Bardeen case
\begin{equation}\label{tau_bardeen}
    \tau_{\mathrm{Bardeen}} = \frac{4\pi r_i\left(a^2+r_i^2\right)}{r_i^2-2a^2}.
\end{equation}
Therefore, the on-shell temperature satisfies
\begin{equation}
    T_{\mathrm{Bardeen}} = \frac{1}{\tau_{\mathrm{Bardeen}}},
\end{equation}
in agreement with Eq.~(\ref{temperature_bardeen}) and with the topological thermodynamic construction~\cite{Wei:2022dzw}.

For the winding numbers and the heat-capacity sign, we use the general intervals given in Eqs.~(\ref{winding_numbers_intervals_general_case}) and~(\ref{heat_capacity_intervals}). For $\alpha=3$ and $\beta=2$, Eq.~(\ref{xcritical_general}) gives the critical value
\begin{equation}
    x_c \approx 2.697.
\end{equation}

For the Bardeen case, the extremal lower bound is $x_{\mathrm{ext}}=\sqrt{2}$. Hence, the physical classification is
\begin{equation}\label{winding_numbers_heat_capacity_intervals_bardeen}
    \begin{cases}
        w_i=+1, \quad \mathrm{Sgn}(C)=+1, & \sqrt{2}<\dfrac{r_i}{a}<2.697, \quad \text{stable branch},\\[0.3cm]
        J_i=0,  & \dfrac{r_i}{a}=2.697, \quad \text{degenerate critical zero},\\[0.3cm]
        w_i=-1, \quad \mathrm{Sgn}(C)=-1, & \dfrac{r_i}{a}>2.697, \quad \text{unstable branch}.
    \end{cases}
\end{equation}
The endpoint $r_{\mathrm{ext}}/a=\sqrt{2}$ is the extremal configuration, where $T=0$ and $\tau\rightarrow\infty$, while $r_c/a=2.697$ is the degenerate critical zero ($J_i=0$) where the heat capacity diverges and changes sign. For $\tau^*>\tau_c^*$, the two regular Bardeen defects lie on opposite sides of this critical radius and carry winding numbers $+1$ and $-1$, respectively.

\subsubsection{Hayward Black Hole for $\alpha = \beta = 3$}

The Hayward black hole is obtained from the generalized metric by choosing $\alpha=\beta=3$. In this case, the mass is
\begin{equation}\label{hayward_mass}
    M_{\mathrm{Hayward}} = \frac{a^3+r_h^3}{2r_h^2}.
\end{equation}
The Hawking temperature is given by
\begin{equation}\label{T_hayward}
    T_{\mathrm{Hayward}} = \frac{r_h^3-2a^3}{4\pi r_h(a^3+r_h^3)}.
\end{equation}
The entropy reads
\begin{equation}\label{entropy_hayward}
    S_{\mathrm{Hayward}} = \frac{\pi\left(r_h^3-2a^3\right)}{r_h}.
\end{equation}
Therefore, the generalized off-shell Helmholtz free energy becomes
\begin{equation}\label{free_energy_hayward}
    \mathcal{F}_{\mathrm{Hayward}} = \frac{a^3(4\pi r_h+\tau) + r_h^3(\tau-2\pi r_h)}{2r_h^2\tau}.
\end{equation}

The zeros of the vector field leads to
\begin{equation}\label{tau_hayward}
    \tau_{\mathrm{Hayward}} = \frac{4\pi r_i\left(a^3+r_i^3\right)}{r_i^3-2a^3}.
\end{equation}
Once again, the on-shell relation
\begin{equation}
    T_{\mathrm{Hayward}}=\frac{1}{\tau_{\mathrm{Hayward}}}
\end{equation}
reproduces the Hawking temperature in Eq.~(\ref{T_hayward}), confirming the consistency of the topological approach.

For $\alpha=\beta=3$, the critical value obtained from the general expression (\ref{xcritical_general}) is
\begin{equation}
    x_c \approx 2.168.
\end{equation}

For the Hayward case, the extremal lower bound is $x_{\mathrm{ext}}=2^{1/3}$. Thus, the winding numbers and the sign of the heat capacity on the physical branch are classified as
\begin{equation}\label{winding_numbers_heat_capacity_intervals_hayward}
    \begin{cases}
        w_i=+1, \quad \mathrm{Sgn}(C)=+1,
        & 2^{1/3}<\dfrac{r_i}{a}<2.168,
        \quad \text{stable branch},\\[0.3cm]
        J_i=0,
        & \dfrac{r_i}{a}=2.168,
        \quad \text{degenerate critical zero},\\[0.3cm]
        w_i=-1, \quad \mathrm{Sgn}(C)=-1,
        & \dfrac{r_i}{a}>2.168,
        \quad \text{unstable branch}.
    \end{cases}
\end{equation}
The endpoint $r_{\mathrm{ext}}/a=2^{1/3}$ is extremal, with $T=0$ and $\tau\rightarrow\infty$, whereas $r_c/a=2.168$ is the degenerate critical zero ($J_i=0$) separating the locally stable and unstable thermodynamic branches. For $\tau^*>\tau_c^*$, the two regular Hayward defects lie on opposite sides of this point and carry winding numbers $+1$ and $-1$.

\subsubsection{Simpson--Visser Case for $\alpha = 1$ and $\beta = 2$}

Finally, we consider the Simpson--Visser case, obtained by setting $\alpha=1$ and $\beta=2$. The mass is given by
\begin{equation}\label{mass_SV}
    M_{\mathrm{SV}} = \frac{1}{2}\sqrt{a^2+r_h^2}.
\end{equation}
The corresponding temperature is
\begin{equation}\label{temperature_SV}
    T_{\mathrm{SV}} = \frac{r_h}{4\pi\left(a^2+r_h^2\right)}.
\end{equation}
The entropy is
\begin{equation}\label{entropy_SV}
    S_{\mathrm{SV}} = \pi\left[r_h\sqrt{a^2+r_h^2} + a^2\tanh^{-1}\left(\frac{r_h}{\sqrt{a^2+r_h^2}}\right)\right].
\end{equation}
From these quantities, the generalized off-shell Helmholtz free energy reads
\begin{equation}\label{free_energy_SV}
    \mathcal{F}_{\mathrm{SV}} = \frac{1}{2}\sqrt{a^2+r_h^2} - \frac{\pi\left[r_h\sqrt{a^2+r_h^2} + a^2 \tanh^{-1}\left(\frac{r_h}{\sqrt{a^2+r_h^2}}\right)\right]}{\tau}.
\end{equation}

The condition for the zeros of the vector field gives
\begin{equation}\label{tau_SV}
    \tau_{\mathrm{SV}} = \frac{4\pi\left(a^2+r_i^2\right)}{r_i}.
\end{equation}
Thus,
\begin{equation}
    T_{\mathrm{SV}}=\frac{1}{\tau_{\mathrm{SV}}},
\end{equation}
which agrees with the Simpson--Visser temperature in Eq.~(\ref{temperature_SV}).

For $\alpha=1$ and $\beta=2$ in (\ref{xcritical_general}), the critical value is
\begin{equation}
    x_c = 1.
\end{equation}

For $\alpha=1$, the extremal expression gives $x_{\mathrm{ext}}=0$. Therefore, unlike the Bardeen and Hayward cases, no positive lower cutoff is introduced, and the physical branch begins at $r_i/a>0$. The winding numbers and the heat-capacity sign are then given by
\begin{equation}\label{winding_numbers_heat_capacity_intervals_SV}
    \begin{cases}
        w_i=+1, \quad \mathrm{Sgn}(C)=+1,
        & 0<\dfrac{r_i}{a}<1,
        \quad \text{stable branch},\\[0.3cm]
        J_i=0,
        & \dfrac{r_i}{a}=1,
        \quad \text{degenerate critical zero},\\[0.3cm]
        w_i=-1, \quad \mathrm{Sgn}(C)=-1,
        & \dfrac{r_i}{a}>1,
        \quad \text{unstable branch}.
    \end{cases}
\end{equation}
The boundary $r_{\mathrm{ext}}/a=0$ corresponds to $T=0$ and $\tau\rightarrow\infty$, while $r_c/a=1$ is the degenerate critical zero ($J_i=0$) at which the stability changes. For $\tau^*>\tau_c^*$, the two regular Simpson--Visser case defects lie on opposite sides of this point and carry winding numbers $+1$ and $-1$.

\section{Results and Discussion}

\begin{figure}[h!]
    \centering
    \includegraphics[width=.70\linewidth]{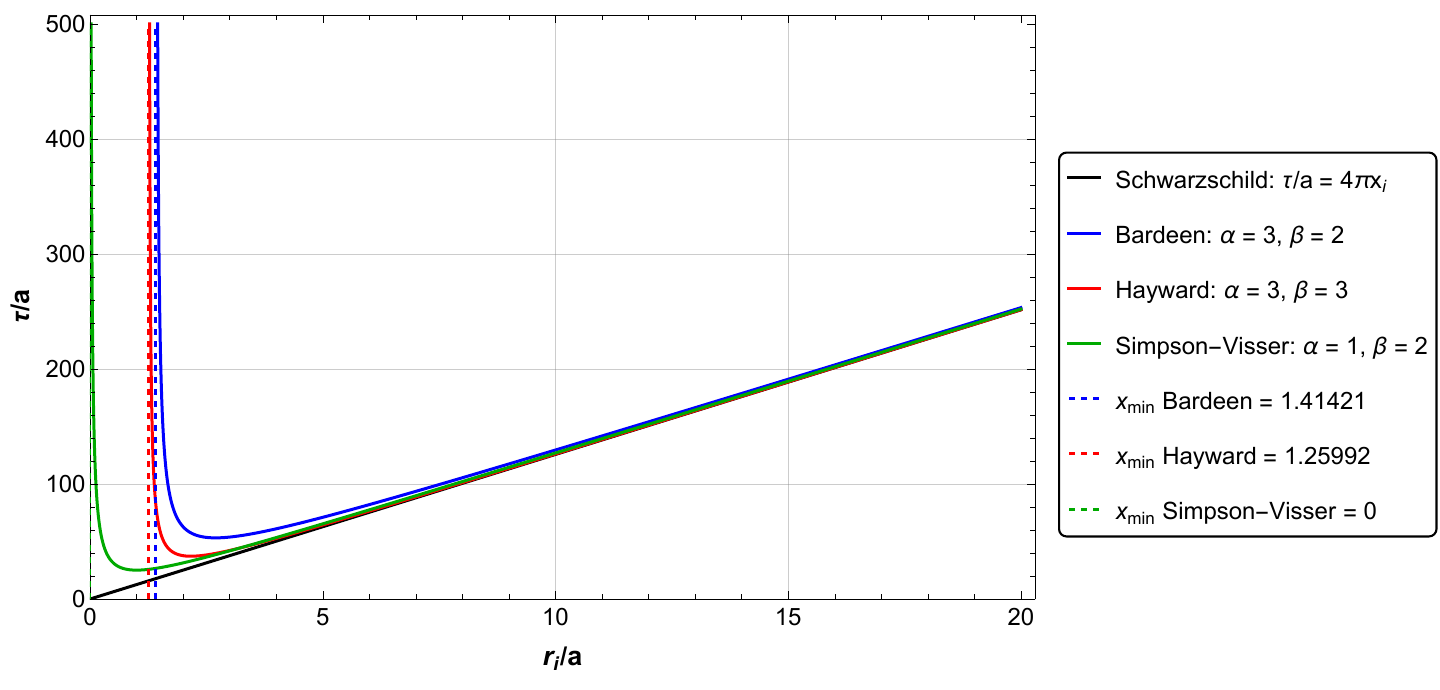}
    \caption{Dimensionless inverse temperature $\tau/a$ as a function of $r_i/a$. The vertical dotted lines mark the lower bounds $x_{\mathrm{ext}}=(\alpha-1)^{1/\beta}$ of the physical outer-horizon branches for $\alpha>1$; for Simpson--Visser-type, $x_{\mathrm{ext}}=0$.}
    \label{fig:Figure 3}
\end{figure}

In the previous section, the topological thermodynamics of the generalized metric given in Eq.~(\ref{2.2}) was investigated. By fixing the parameters $\alpha$ and $\beta$, the Bardeen, Hayward, and Simpson--Visser-type geometries are recovered as particular cases. Starting from the general formulation, it was possible to obtain the relevant topological quantities and to verify the consistency of the approach by recovering the Hawking temperature from the on-shell relation $\tau = 1/T$, evaluated at $r_h = r_i$. Figures~\ref{fig:Figure 4} and~\ref{fig:Figure 3} show, respectively, the dimensionless temperature $T^* = aT$ and the dimensionless inverse temperature $\tau^* = \tau/a$ as functions of the dimensionless horizon radius $x_i = r_i/a$.

All thermodynamic-topology results discussed below refer to the fixed-$a$ ensemble specified in Sec.~III, with $\alpha$ and $\beta$ treated as fixed spacetime parameters. Accordingly, the comparison among the different geometries concerns different fixed parameter choices within the same thermodynamic prescription, rather than variations of $a$, $\alpha$, or $\beta$ as thermodynamic variables.

Figure~\ref{fig:Figure 3} shows the behavior of $\tau/a$ as a function of $r_i/a$. For $\alpha>1$, the vertical dotted lines indicate the extremal radii $x_{\mathrm{ext}}=r_{\mathrm{ext}}/a=(\alpha-1)^{1/\beta}$, which define the lower bounds of the physical outer-horizon branches. At these endpoints, the Hawking temperature vanishes and $\tau/a\rightarrow\infty$. Accordingly, the Bardeen and Hayward branches must be restricted to $r_i/a>\sqrt{2}$ and $r_i/a>2^{1/3}$, respectively. In the Simpson--Visser case, $\alpha=1$ and the lower bound becomes $x_{\mathrm{ext}}=0$, so the physical branch is $r_i/a>0$. In every case, the stable segment is the portion of the physical branch lying below $x_c$, not the full algebraic interval $0<r_i/a<x_c$. For sufficiently large $r_i/a$, all finite-$a$ cases asymptotically recover the Schwarzschild behavior.

The critical values of $\tau/a$ for each spacetime are summarized in Table~\ref{tab:critical_tau_values}. These critical points correspond to the points where the heat capacity diverges and changes sign. Thus, they separate locally stable and unstable thermodynamic branches. In contrast, the Schwarzschild case does not exhibit a finite critical point, since $\tau/a = 4\pi r_i/a$ is a monotonic function.

\begin{table}[h!]
\centering
\begin{tabular}{|c|c|c|}
\hline
\textbf{Spacetime}                         & \textbf{$r_i/a$}                      & \textbf{$\tau/a$}            \\ \hline
Bardeen ($\alpha = 3$, $\beta = 2$)        & 2.69721                               & 53.1706                      \\ \hline
Hayward ($\alpha = \beta = 3$)             & 2.16843                               & 37.2232                      \\ \hline
Simpson--Visser ($\alpha = 1$, $\beta = 2$) & 1                                     & 25.1327                      \\ \hline
Schwarzschild (asymptotic case)            & No critical point                     & No critical point            \\ \hline
\end{tabular}
\caption{Critical values of $\tau/a$ as a function of $r_i/a$.}
\label{tab:critical_tau_values}
\end{table}

The values listed in Table~\ref{tab:critical_tau_values} are the minima $\tau_c^*$ of the corresponding inverse-temperature curves. Therefore, the pair of regular defects exists only for $\tau^*>\tau_c^*$. At $\tau^*=\tau_c^*$, the two defects coalesce at $x=x_c$ into a degenerate zero with $J=0$, whereas for $\tau^*<\tau_c^*$ no physical zero exists on the corresponding finite-$a$ branch. Thus, the critical point is interpreted as a defect-pair coalescence point rather than as an isolated defect with winding number $w=0$.

Since $T^* = aT = 1/\tau^*$, the critical values of the dimensionless temperature are the inverse of the corresponding critical values of $\tau/a$. These values are presented in Table~\ref{tab:critical_aT_values}. The Simpson--Visser case reaches the highest maximum temperature, followed by the Hayward and Bardeen cases. Equivalently, the minimum value of $\tau/a$ is lowest for Simpson--Visser case and highest for Bardeen. This behavior shows that the parameters $\alpha$ and $\beta$, which encode the regularization mechanism of the spacetime, directly affect the location of the critical points and the thermodynamic stability structure.

\begin{table}[h!]
\centering
\begin{tabular}{|c|c|c|}
\hline
\textbf{Spacetime}                         & \textbf{$r_i/a$}             & \textbf{$aT$}                \\ \hline
Bardeen ($\alpha = 3$, $\beta = 2$)        & 2.69721                      & 0.0188074                    \\ \hline
Hayward ($\alpha = \beta = 3$)             & 2.16843                      & 0.0268649                    \\ \hline
Simpson--Visser case ($\alpha = 1$, $\beta = 2$) & 1                            & 0.0397887                    \\ \hline
Schwarzschild (asymptotic case)            & No critical point            & No critical point            \\ \hline
\end{tabular}
\caption{Critical values of $aT$ as a function of $r_i/a$.}
\label{tab:critical_aT_values}
\end{table}

It is important to distinguish the parameter dependence of the critical quantities from the global thermodynamic topological classification. The values of $x_c$, $\tau_c$, and $T_c$ vary with $\alpha$ and $\beta$, showing that these parameters control the positions of the defects and the extent of the locally stable branch. Nevertheless, these continuous shifts do not change the total topological number of the three finite-$a$ cases considered. Whenever two regular defects exist, their winding numbers are opposite and their sum remains $W=0$. Thus, the Bardeen, Hayward, and Simpson--Visser case geometries share the same global thermodynamic topological class in the present setup, although their local thermodynamic structures are quantitatively different. The Schwarzschild case is qualitatively distinct because its inverse-temperature curve is monotonic and supports only one unstable defect, yielding $W=-1$. Accordingly, the novelty of the present analysis lies in the family-wide analytical criterion that separates parameter-dependent local stability data from the global topological class, rather than in the proposal of a previously unknown value of $W$.

\begin{figure}[htbp]
    \centering
    \includegraphics[width=0.95\linewidth]{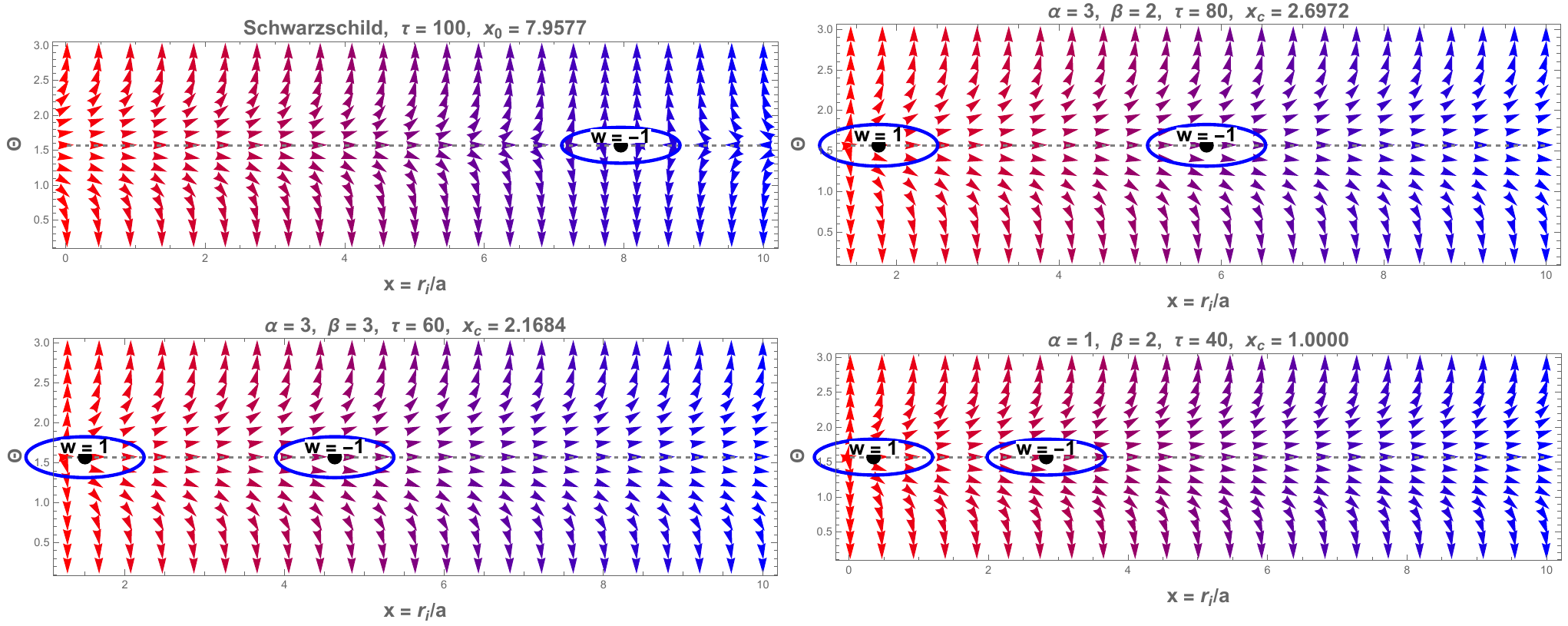}
    \caption{Normalized vector field $\vec{n}=\vec{\phi}/|\vec{\phi}|$ for the Schwarzschild, Bardeen, Hayward, and Simpson--Visser cases. The circles indicate the zeros of the vector field, and the corresponding winding numbers are shown near each defect.}
    \label{fig:Figure 5}
\end{figure}

To complement the analysis based on the extrema of $\tau/a$ and $aT$, Fig.~\ref{fig:Figure 5} displays the normalized vector field
\begin{equation}
    \vec{n}=\frac{\vec{\phi}}{|\vec{\phi}|},
\end{equation}
where
\begin{equation}
    \vec{\phi}=\left(\frac{\partial \mathcal{F}}{\partial r_h},-\csc\Theta\cot\Theta\right).
\end{equation}
The zeros of this vector field occur when both components vanish simultaneously. The second component vanishes at $\Theta=\pi/2$, while the first component gives the condition $\partial \mathcal{F}/\partial r_h=0$. Therefore, the defects appear along the line $\Theta=\pi/2$ and are associated with the values of $r_i/a$ that satisfy the off-shell condition for a fixed value of $\tau$.

For the Schwarzschild case, the off-shell free energy leads to a single zero for a fixed value of $\tau$. Since the heat capacity of the Schwarzschild black hole is always negative, this defect carries winding number $w=-1$, in agreement with its thermodynamic instability. This is consistent with the absence of a finite critical point in Tables~\ref{tab:critical_tau_values} and~\ref{tab:critical_aT_values}, because $\tau/a=4\pi r_i/a$ is monotonic and does not present a local minimum.

For the regular cases, the behavior is different. The Bardeen, Hayward, and Simpson--Visser case spacetimes exhibit two regular zeros of the vector field for the selected values of $\tau$, which lie above the corresponding critical inverse temperatures. These zeros lie on opposite sides of the critical value in equation (\ref{winding_numbers_intervals_general_case})
\begin{equation}
    x_c=\left[\frac{\alpha(\beta+1)-2+\sqrt{[\alpha(\beta+1)-2]^2+4(\alpha-1)}}{2}\right]^{1/\beta}.
\end{equation}
According to the sign of $\left.\frac{\partial^2\mathcal{F}}{\partial r_h^2}\right|_{r_h=r_i}$, the defect located in the physical interval $x_{\mathrm{ext}}<r_i/a<x_c$ has winding number $w=+1$, whereas the defect located in the region $r_i/a>x_c$ has winding number $w=-1$. Thus, the first defect corresponds to a locally stable outer-horizon branch, while the second corresponds to an unstable branch. The region $0<r_i/a\leq x_{\mathrm{ext}}$ is excluded from this thermodynamic classification for $\alpha>1$.

In the Bardeen solution, with $(\alpha,\beta)=(3,2)$, the critical value is $x_c=2.6972$. For the value $\tau=80$ used in Fig.~\ref{fig:Figure 5}, the vector field displays two defects: one on each side of $x_c$. The left defect has $w=+1$, while the right defect has $w=-1$. The same qualitative structure occurs for the Hayward solution, where $(\alpha,\beta)=(3,3)$ and $x_c=2.1684$, and for the Simpson--Visser case, where $(\alpha,\beta)=(1,2)$ and $x_c=1$. In all these regular cases, the two defects appear as a pair with opposite winding numbers.

Therefore, for the regular spacetimes analyzed here, the total topological charge is
\begin{equation}
    W=\sum_i w_i = +1-1=0.
\end{equation}
This result shows that, for $\tau^*>\tau_c^*$, the regular black-hole configurations considered in this work contain one stable and one unstable defect whose charges cancel. As $\tau^*$ decreases toward $\tau_c^*$, these defects approach each other and coalesce at $x_c$, where $J=0$ and the regular-zero winding formula no longer applies. For $\tau^*<\tau_c^*$, the pair is absent. The heat-capacity divergence and sign change occur at the same critical radius, providing a direct thermodynamic counterpart of the defect-pair coalescence observed in the vector-field description.

Overall, the results confirm the consistency between the topological approach and the standard thermodynamic analysis. The zeros of the vector field reproduce the inverse Hawking temperature through the on-shell condition $\tau=1/T$, while the sign of the winding number agrees with the sign of the heat capacity. The Schwarzschild case exhibits a single unstable defect with $w=-1$, reflecting its negative heat capacity. In contrast, the Bardeen, Hayward, and Simpson--Visser cases exhibit pairs of defects with opposite winding numbers, $w=+1$ and $w=-1$, corresponding respectively to locally stable and unstable thermodynamic branches. Thus, the generalized Bardeen metric provides a unified framework in which different regular black holes can be compared through their topological thermodynamic structure.

The present result can also be placed within the universal thermodynamic topological classification introduced in Ref.~\cite{Wei:2024gfz}. In the finite-$a$ cases considered here, the smaller-radius regular defect belongs to the locally stable branch and carries winding number $w=+1$, whereas the larger-radius regular defect belongs to the unstable branch and carries $w=-1$. Therefore, the ordering of the regular defects is $[+,-]$. Together with the inverse-temperature behavior discussed above, this identifies the Bardeen, Hayward, and Simpson--Visser cases, within the fixed-$a$ ensemble adopted here, with the previously established $W^{0+}$ thermodynamic topological class. Thus, the value $W=0$ obtained in the present work does not define a new global thermodynamic topological class.

It is also useful to compare the present results with the recent analysis of regular black holes constructed from pure gravity \cite{WANG2025139402}. In that work, regular black holes arise from an infinite tower of higher-curvature corrections in quasi-topological gravity for $D\geq5$, without requiring a specific matter source for regularization. This construction is therefore physically distinct from the four-dimensional Neves--Saa family considered here, where regularity is encoded through the metric parameters and the effective geometric structure.

Despite these differences, both approaches lead to the same global thermodynamic topological classification. The higher-curvature regular black holes of Ref.~\cite{WANG2025139402} and the Bardeen, Hayward, and Simpson--Visser cases analyzed here belong to the $W^{0+}$ class, characterized by the defect ordering $[+,-]$, with $w_{\mathrm{small}}=+1$ and $w_{\mathrm{large}}=-1$. Nevertheless, the two analyses differ in their gravitational origin, dimensional setting, and thermodynamic prescription. While Ref.~\cite{WANG2025139402} employs Wald entropy within a higher-curvature gravity framework, the present work adopts the fixed-$a$ ensemble with the entropy determined from $dM=T\,dS$, keeping $a$, $\alpha$, and $\beta$ fixed. Therefore, the agreement in the $W^{0+}$ classification emerges despite the distinct regularization mechanisms and thermodynamic constructions.

\section{Conclusion}

In this work, we investigated the topological thermodynamics of the generalized Bardeen black hole through the off-shell Helmholtz free energy formalism.

Here, the term generalized Bardeen black hole refers to the Neves--Saa family, from which the Bardeen, Hayward, and Simpson--Visser-type geometries are recovered as particular limiting cases. Therefore, the conclusions obtained below concern the thermodynamic-topological structure of the full family and its regular black-hole limits, rather than only the original Bardeen realization.

This geometry provides a unified framework that contains the Bardeen, Hayward, and Simpson--Visser-type spacetimes as particular cases, obtained by suitable choices of the parameters $\alpha$ and $\beta$. By analyzing the zeros of the vector field
\begin{equation}
    \vec{\phi}=\left(\frac{\partial \mathcal{F}}{\partial r_h},-\csc\Theta\cot\Theta\right),
\end{equation}
we showed that the topological approach consistently reproduces the standard thermodynamic behavior of these black holes.

The condition $\partial\mathcal{F}/\partial r_h=0$ leads to the on-shell relation $\tau=1/T$, confirming that the zeros of the vector field are directly associated with the thermodynamic states of the black hole. In dimensionless form, the inverse temperature $\tau^*=\tau/a$ and the temperature $T^*=aT$ were analyzed for the Schwarzschild, Bardeen, Hayward, and Simpson--Visser cases. The regular black holes exhibit an extremal radius at which the Hawking temperature vanishes, while their temperature curves present a finite maximum. This maximum corresponds to the points where the heat capacity diverges and changes sign.

The winding-number analysis shows that the Bardeen, Hayward, and Simpson--Visser cases do not constitute distinct global thermodynamic topological classes within the setup considered here. For the selected values of $\tau$ considered here for which two regular zeros exist, the defect located on the smaller-radius physical branch, $x_{\mathrm{ext}}<r_i/a<x_c$, has winding number $w=+1$ and corresponds to a locally stable configuration, whereas the defect on the larger-radius branch, $r_i/a>x_c$, has winding number $w=-1$ and corresponds to an unstable configuration. Their contributions cancel, so that $W=\sum_i w_i = +1-1=0$. 

The common value $W=0$ shows that varying $\alpha$ and $\beta$ among the three finite-$a$ cases analyzed does not, by itself, generate a different global topological class. Instead, these parameters determine the extremal lower bound, the critical radius $x_c$, the critical Euclidean period $\tau_c$, the corresponding maximum temperature, and the extent of the locally stable branch. They therefore modify the local and quantitative thermodynamic structure of the defect pair. For $\tau^*>\tau_c^*$ the pair consists of two regular defects with charges $+1$ and $-1$; at $\tau^*=\tau_c^*$ they coalesce into a degenerate zero with $J=0$, to which no individual winding number $w=0$ is assigned; and for $\tau^*<\tau_c^*$ no physical defect remains on the finite-$a$ branch.

Within the universal thermodynamic topological classification of Ref.~\cite{Wei:2024gfz}, the stable-to-unstable ordering $[+,-]$ of the two regular defects places the Bardeen, Hayward, and Simpson--Visser cases in the established $W^{0+}$ class. Therefore, the common value $W=0$ obtained here should be understood as a realization of an already known global class rather than as a new thermodynamic topological category. The Schwarzschild limit is correspondingly consistent with the $W^{1-}$ class. This comparison reinforces that the specific contribution of the present analysis is the analytical characterization of how $\alpha$ and $\beta$ shift the local thermodynamic and critical data while preserving the same global $W^{0+}$ structure for the three finite-$a$ cases considered.

The Schwarzschild limit is qualitatively different because its inverse-temperature curve is monotonic and contains only one unstable defect with $w=-1$, giving $W=-1$. The principal contribution of this work is therefore the analytical, family-wide separation between parameter-dependent local thermodynamic information and the global thermodynamic topological class. We do not claim that $W=0$ is a previously unknown class; rather, we show explicitly how the Neves--Saa parameters move the critical points and alter the stability domains while the Bardeen, Hayward, and Simpson--Visser cases considered remain in the same global class. These conclusions refer specifically to the fixed-$a$ ensemble adopted throughout the analysis; allowing $a$ to vary would require an extended first law and, in general, an ensemble-dependent off-shell thermodynamic potential. Future investigations may extend this analysis to charged, rotating, or higher-dimensional generalizations, as well as to ensembles in which additional thermodynamic variables can modify the topological structure of the phase space.

\acknowledgments  
\hspace{0.5cm} The authors thank the Coordena\c{c}\~{a}o de Aperfei\c{c}oamento de Pessoal de N\'{i}vel Superior (CAPES) and the Conselho Nacional de Desenvolvimento Cient\'{i}fico e Tecnol\'ogico (CNPq).

\bibliographystyle{apsrev4-1}
\bibliography{ref}

\end{document}